\documentclass[prb,twocolumn]{revtex4} 
\usepackage{graphicx,subfigure,dcolumn,bm}  

\begin{document}

\title{Bulk Aluminum at High Pressures: A First-Principles Study}

\author{Michael J. Tambe}
\author{Nicola Bonini}
\author{Nicola Marzari}
\email{marzari@mit.edu}
\affiliation{Department of Materials Science and Engineering, Massachusetts Institute of Technology, Cambridge MA 02139}

\begin{abstract}
  The behavior of metals at high pressure is of great importance to
  the fields of shock physics, geophysics, astrophysics, and nuclear
  materials.  We study here bulk crystalline
  aluminum from first-principles at pressures up to 2500 GPa -
  soon within reach of laser-based experimental facilities. Our simulations
  use density-functional theory and density-functional perturbation
  theory in the local-density and generalized-gradient
  approximations.  
  Notably, the two
different 
  exchange-correlation functionals predict very similar results for
  the $fcc \rightarrow hcp$, $fcc \rightarrow bcc$, and $hcp
  \rightarrow bcc$ transition pressures, around $175$ GPa, $275$ GPa,
  and $380$ GPa respectively. 
In addition, our results indicate that core overlaps become noticeable
  only beyond pressures of 1200 GPa. 
From the phonon dispersions of the fcc
  phase at increasing pressure, we predict a softening of the lowest
  transverse acoustic vibrational mode along the [110] direction,
  which corresponds to a Born instability of the fcc phase around
  $725$ GPa.
\end{abstract}

\maketitle

First-principles calculations have proved useful to the fields of
geophysics,~\cite{Anderson.RevGeoPhys.30.57}
astrophysics,~\cite{Alavi.Science.269.1252} and nuclear
materials.~\cite{Cheeseman.JCP.104.5497} 
Aluminum, being cubic close-packed and having no d-shell electrons, is
a prototype for theoretical predictions and understanding the
high-pressure behavior of simple metals.~\cite{Martin.Nature.400.117}
Currently the National Ignition Facility~\cite{NIF} at LLNL
is expected to achieve shockless
compression~\cite{PhysRevLett.92.075002} of metals up to 2000 GPa.
This new facility may provide rapid advancements to
high-pressure physics and could partner very successfully
with theoretical studies.

The equation of state (EOS)
and phase stability of aluminum were first studied from first-principles in the early
1980s.~\cite{PhysRevB.29.6434,PhysRevB.27.5986,PhysRevB.27.3235}  In
all cases the predicted phase sequence was $fcc \rightarrow hcp
\rightarrow bcc$, but predictions differed greatly in the transition
pressures. Several other calculations
within the local-density approximation (LDA)~\cite{PhysRevB.53.3007}
or the generalized-gradient approximation (GGA)~\cite{JPhysCondMatt.14.6989,JPhysCondMatt.18.10881}
have since then been performed, with a predicted static (i.e. without the phonon contribution) $fcc \rightarrow hcp$ 
transition pressure of $205 \pm 20$ GPa~\cite{PhysRevB.53.3007} in LDA and
170 GPa~\cite{JPhysCondMatt.14.6989}
and 192 GPa~\cite{JPhysCondMatt.18.10881} in GGA.
These discrepancies are more notable for the $hcp \rightarrow bcc$ transition pressure:
$565 \pm 60$ GPa~\cite{PhysRevB.53.3007} in LDA versus 360 GPa~\cite{JPhysCondMatt.14.6989} in GGA, leaving
significant uncertainties open.
Theoretical work on the vibrational properties of
aluminum also suggests for the $fcc \rightarrow hcp$ transition a transition pressure
higher than the static one.~\cite{JPhysCondMatt.14.6989,JPhysCondMatt.18.10881} 
Elastic properties~\cite{JPhysCondMatt.10.9889,PhysRevLett.82.3296}
and the absolute strength under 
tension~\cite{PhysRevLett.91.135501} have also been calculated;
the latter results are of particular interest as they demonstrate the important role vibrational 
modes play in determining mechanical stability and suggest that shear failure modes 
are inherent in aluminum.

Experimentally, the equation of state of aluminum at high pressures
was studied by shock-compression~\cite{PhysRevLett.60.1414}
at pressures above the predicted 
maximum for the $fcc \rightarrow hcp$ phase boundary,~\cite{PhysRevB.53.3007}
but a transition was not observed. However recent diamond anvil cell experiments
observed a $fcc \rightarrow hcp$ transition at $217 \pm
10$ GPa~\cite{PhysRevLett.96.045505} highlighting the difficulty in
achieving thermodynamic equilibrium in shock-compression.

In this article we report first-principles calculations of aluminum
under hydrostatic compression up to $2500$ GPa.  In order to assess mechanical
stability under shock, we also calculate the vibrational properties in
the fcc phase and determine the elastic constants from the slopes of
the phonon dispersions (i.e. the sound velocities).  

The equations of state in the fcc, bcc, and hcp phases have been
calculated with density-functional theory (DFT) 
within both LDA~\cite{PhysRevB.23.5048} and
GGA.~\cite{PhysRevLett.77.3865}  Calculations have used the 
{\sc Quantum-ESPRESSO} package.~\cite{PWSCF.PP} We use plane-wave basis sets
and pseudopotentials and both 3 electron (3e) 
norm-conserving pseudopotentials,~\cite{PhysRevB.41.1227} with the 3s and 3p
electrons in the valence and nonlinear core-corrections, and
11 electron (11e) ultrasoft pseudopotentials~\cite{PhysRevB.41.7892}
where the 2s and 2p electrons, usually frozen in the core, are
included explicitly in the valence.  The inclusion of 
the 2s and 2p electrons in the  valence is essential to investigate the 
relevance of inner core electrons at high pressure. The planewave
cutoffs for the wavefunctions are 25 Ry and 100 Ry for
the 3e and 11e pseudopotentials, respectively, and 150 Ry and 800 Ry for the
charge density.  Brillouin zone integrations have been
performed using a cold smearing~\cite{PhysRevLett.82.3296} of
0.02 Ry over shifted Monkhorst-Pack meshes of order $16 \times 16
\times 16$ for the fcc, $22 \times 22 \times 22$ for the bcc and $16
\times 16 \times 10$ for the hcp phases. The large 
sizes of the k-point meshes are necessary to obtain 
fully-converged transition pressures.
The data for the total energy as function of volume have been fitted to the
Birch 3rd order EOS~\cite{Birch.PhysRev.71.809} near equilibrium to obtain equilibrium
volumes and bulk moduli. The data for the equation of state have been determined from calculations performed at around
$50 - 100$ different volumes over a pressure range of $0 - 2500$ GPa.
Finally, the vibrational properties have been computed using density-functional
perturbation theory (DFPT).~\cite{RevModPhys.73.515}  The dynamical
matrices have been calculated on a $4 \times 4 \times 4$ $q$-point
mesh and Fourier interpolation has been used to evaluate the phonon
frequencies on finer grids.

\begin{table}
	\caption{The equilibrium lattice parameters, bulk moduli, and transition pressures of Al calculated with the different pseudopotentials described in the text, and compared to experimental results.}
	\begin{center}
		\begin{tabular}{|llllll|}
			\hline
			& $a_0$ [\AA]& $B_0$ [GPa] & $fcc \rightarrow hcp$ & $fcc \rightarrow bcc$ & $hcp \rightarrow bcc$ \\
			\hline
			11e GGA & 4.044 & 73.2 & 175 GPa & 275 GPa &  383 GPa \\
			\hline
			11e LDA & 3.985 & 83.7 & 172 GPa & 272 GPa & 380 GPa \\
			\hline
			3e GGA & 4.055 & 71.8 & 180 GPa & 285 GPa & 420 GPa \\
			\hline
			Expt. & 4.025$^a$ & 78.3$^a$ & 217 GPa$^b$ & --- & --- \\
			\hline
		\end{tabular}
	\end{center}
	\flushleft$^a$ Ref.~\onlinecite{Stedman.PhysRev.145.492}\\
	$^b$ Ref.~\onlinecite{PhysRevLett.96.045505}
	\label{EOScompPP}
\end{table}

\begin{figure}
	\begin{center}
		\includegraphics[width=80mm]{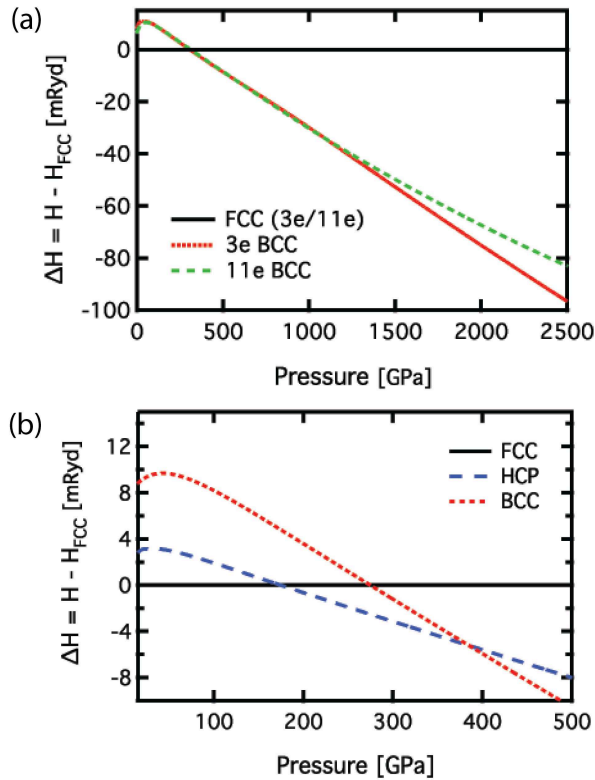}
	\end{center}
	\caption{(a) Enthalpy of the bcc phase (3e and 11e GGA) 
relative to the enthalpy of the fcc phase (3e and 11e GGA, respectively). (b) 
Enthalpies for hcp and bcc phase relative to the enthalpy of 
the fcc phase. The $fcc \rightarrow hcp$, $fcc \rightarrow bcc$, 
and $hcp \rightarrow bcc$ transition pressures are 
$175$ GPa, $275$ GPa, and $380$ GPa respectively.}
	\label{EOS}
\end{figure}

The role of the inner core electrons is of primary concern at very
high pressures.  Under normal conditions, there is not sufficient
overlap between the core and valence shell electrons to question 
the frozen-core approximation,~\cite{PhysRevB.12.2111} but
at the pressures considered here core overlaps may become significant.
To study the validity of the frozen-core approximation, we first
compared the equations of state for 
different phases using both the 
3e and 11e pseudopotentials. We report in Table~\ref{EOScompPP}
the equilibrium lattice parameters and bulk moduli in the
fcc phase at zero pressure, and
in Fig.~\ref{EOS}a the relative enthalpies of fcc and
bcc Al with respect to the fcc phase,
up to 2500 GPa.  These results show that 1) there is little difference between the LDA
and GGA predictions, hinting 
at a broad applicability of density-functional theory in either
approximation, and 2) that the role of the core electrons 
starts to become noticeable only around 1200
GPa, even if already at zero pressure the cores of the 3e pseudopotential start to overlap.~\cite{note_pseudo} 
The equations of state for aluminum in 
the fcc, bcc and hcp phase, using the 11e GGA
pseudopotentials, are shown
in Fig.~\ref{EOS}b.
Although the 11e and 3e calculations give consistent results
up to 1200 GPa, the calculated
transition pressures can vary, particularly for the $hcp \rightarrow bcc$ transition.  This
could easily derive from the fact that enthalpy differences
between these three phases are only a few mRy (see Fig.~\ref{EOS}b), and so,
even at full computational convergence of all parameters, small effects (e.g.
core-state relaxations), which could shift the calculated enthalpy by less than a mRy,
can significantly affect the calculated transition pressures. On the other hand, core electrons seem to
have a negligible effect in determining 
the equilibrium volume, bulk modulus, and even phonon dispersions (see below).

\begin{figure}
	\begin{center}
		\includegraphics[width=80mm]{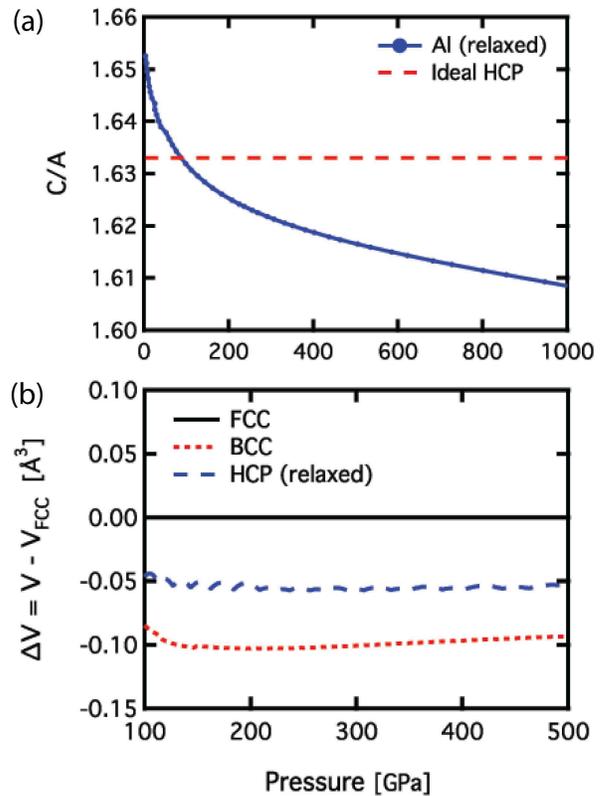}
	\end{center}
	\caption{(a) The c/a ratio of the hcp phase plotted with increasing pressure. (b) Volume versus pressure for the fcc, hcp, and bcc phases plotted relative to the fcc phase.}
	\label{CA-PV}
\end{figure}

Our LDA and GGA
results are consistent with 
previously reported GGA calculations;~\cite{JPhysCondMatt.14.6989,JPhysCondMatt.18.10881}
discrepancies arise with the LDA results reported in Ref.~\onlinecite{PhysRevB.53.3007}, 
that predict $205 \pm 20$ GPa and $565 \pm 60$ GPa for $fcc \rightarrow hcp$ and
$hcp \rightarrow bcc$ transition pressures, respectively.  This discrepancy could arise from 
Ref.~\onlinecite{PhysRevB.53.3007} using only $10 - 15$ points to fit the equation of state:
as reported there, this approximation could
significantly affect transition pressures due to the aforementioned small enthalpy differences 
between competing structures.  We also observe that all parameters of the calculation, 
and particularly the k-point sampling of the Brillouin zone, need to be carefully converged.

Although the pressure that we obtain for the $fcc
\rightarrow hcp$ transition, 175 GPa, is consistent with previous works~\cite{JPhysCondMatt.14.6989,JPhysCondMatt.18.10881}, 
this result is lower than the experimental value of 217 GPa.  As suggested in 
Refs. \onlinecite{JPhysCondMatt.14.6989} and \onlinecite{JPhysCondMatt.18.10881} this discrepancy 
could arise from excluding the phonon contribution to the free energy -
a hypothesis that should be thoroughly tested, but that is
beyond the scope of this brief report.

\begin{figure}
	\begin{center}
		\includegraphics[width=80mm]{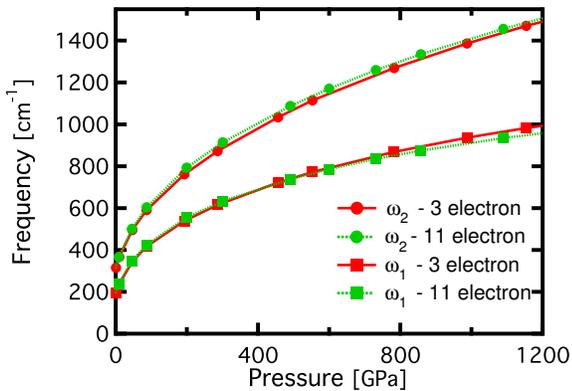}
	\end{center}
	\caption{ The frequencies of the two non-degenerate acoustic phonon modes at X calculated with both the 3e and 11e pseudopotentials and plotted as a function of increasing pressure.}
	\label{FreqP}
\end{figure}

\begin{figure}
	\begin{center}
		\includegraphics[width=80mm]{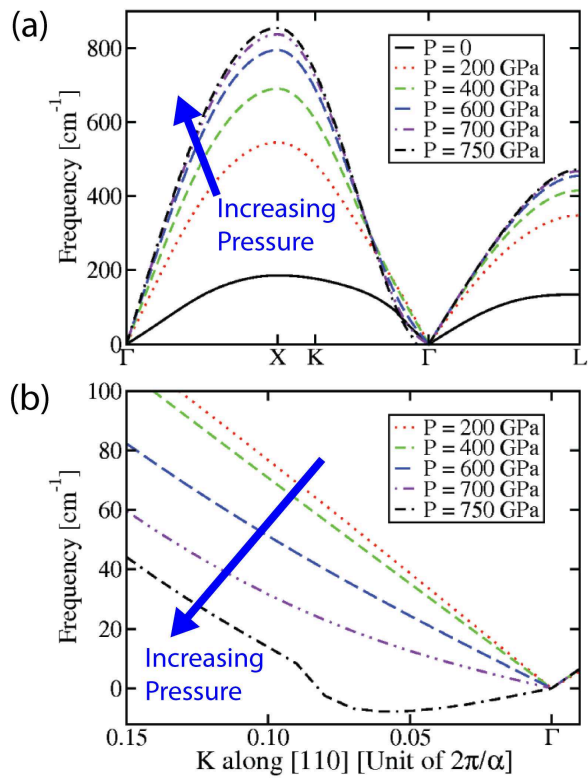}
	\end{center}
	\caption{ (a) The lowest energy branch of phonon dispersion for fcc Al with increasing pressure. (b) As above, enlarged around $\Gamma$, in the K $\rightarrow \Gamma$ direction. A steady flattening is observed with increasing pressure.}
	\label{phonons}
\end{figure}

It should be noted that in our simulations the hcp phase was always fully relaxed
to identify the optimal, equilibrium $c/a$ ratio; this is shown in Fig.~\ref{CA-PV}a.  
Comparison with experiment at $292$ GPa finds agreement
in the c/a ratio to within $0.1\%$ and well within experimental
uncertainty.~\cite{PhysRevLett.96.045505}  At $222$ GPa the
predicted value of the c/a ratio differs from experiment by $1\%$.
Since experiments observe a region between $217$ and $260$ GPa
in which the fcc and hcp phases coexist, this discrepancy is
made more reasonable considering that the system might be out of 
equilibrium. We also show in Fig.~\ref{CA-PV}b the equilibrium volumes 
for the different phases as a function
of pressure. Volume differences between the
phases are -0.055, -0.104, and -0.040 $\AA^{3}$ for the $fcc
\rightarrow hcp$, $fcc \rightarrow bcc$, and $hcp \rightarrow bcc$
transitions respectively, corresponding to volume changes of
$0.6(6)\%$, $1.4(0)\%$, and $0.6(1)\%$.
We note that in our phase sequence, and in those
discussed in the literature,~\cite{PhysRevB.29.6434,PhysRevB.27.5986,PhysRevB.27.3235}
only the fcc, hcp, and bcc phases are considered. As a brief self-check we
performed variable-cell relaxations at 1000 GPa using a four-atom
unit-cell and five random distinct perturbations.  The bcc structure
was always found.

In order to estimate the dynamical response of aluminum
under compression we calculated both the phonon dispersions of the
fcc phase and the cubic elastic constants (these were derived from the sound velocities, i.e.
the slope of the phonons dispersions around $\Gamma$) as a function of
pressure. Experiments have shown that the fcc
phase may exist at pressures above the transition pressure either as a
super-pressurized phase~\cite{PhysRevLett.60.1414} or as a two phase
region.~\cite{PhysRevLett.96.045505} Therefore the mechanical
properties of fcc aluminum at pressures beyond the equilibrium
transition pressure, and any mechanical instabilities that
may lead to mechanical failure are relevant to high pressure
experiments.

\begin{figure}
	\begin{center}
		\includegraphics[width=80mm]{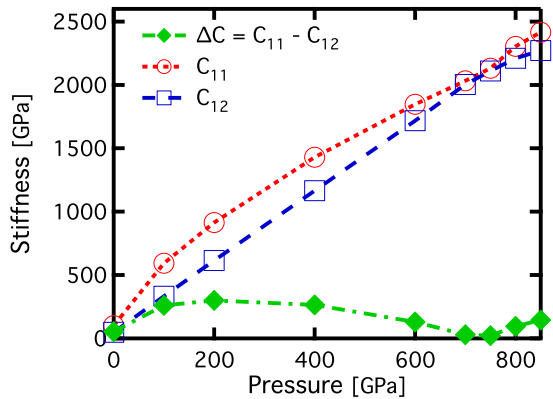}
	\end{center}
	\caption{ Elastic constants as obtained from the sound velocities along the [100], 
	  [110], and [111] directions. According to the Born criterion the fcc phase becomes
	  mechanically unstable when $\Delta c = c_{11} - c_{12} = 0$; this occurs around 725 GPa.}
	\label{elast}
\end{figure}

We calculated the phonon dispersions up to 1150 GPa using DFPT. 
For aluminum this method has been shown to accurately
reproduce experimental values at P=0.~\cite{RevModPhys.73.515}  Our calculations
were performed with the
3e GGA pseudopotential, but compared at selected points in the Brillouin zone
with 11e GGA calculations at pressures up to 1200 GPa. As shown in figure~\ref{FreqP}
the discrepancies between the 3e and 11e results are at most of the order of 3-4\% at
the highest pressure, and much smaller below that.
The phonon dispersions are shown in Fig.~\ref{phonons}a and we highlight the appearance, 
with increasing pressure, of a distinct softening
of the lowest energy mode in the $[110]$ direction.  This 
starts to become evident at approximately $400$ GPa, and is complete at $725$ GPa, as
highlighted in Fig.~\ref{phonons}b. Since the slope of the dispersion curves is
directly related to the elastic constants (Eq. \ref{born-criteria}), 
we can extract the stiffness tensor from the vibrational modes
near $\Gamma$; in our case the Born~\cite{ProcCamPhilSoc.36.160} criterion for stability is

\begin{equation}
	\frac{1}{2}m 
	\left(
		\frac{1}{\hbar} \frac{\partial E}{\partial k_{110}} 
	\right
	) ^2 = c_{11} - c_{12} \geq 0\;.
	\label{born-criteria}
\end{equation}

As Fig.~\ref{elast} shows, the stiffness against shear deformation, $\Delta c = c_{11} - c_{12}$, decreases above 
$400$ GPa and goes to zero around $725$ GPa, resulting in a Born~\cite{ProcCamPhilSoc.36.160} instability. 
These results complement existing studies of the properties of bulk aluminum~\cite{JPhysCondMatt.14.6989,JPhysCondMatt.10.9889} and 
suggest another shear failure mode, supporting previous studies suggesting shear failure 
modes to be inherent to bulk aluminum.~\cite{JPhysCondMatt.10.9889,PhysRevLett.91.135501} More advanced 
treatments of mechanical stability including such effects of anharmonic modes~\cite{RevModPhys.73.515} 
and internal shear stresses created by loading~\cite{PhysRevB.15.3087,PhysRevB.52.12627,Yip.HMM.2005} need 
to be considered in relation to the specific experimental setup before reliable maximum stable pressures can be definitively determined.

Funding for this project has been provided by the U.S. Department of Energy Contract No. DE-FG02-05ER46253. The authors would like to thank A. Dal Corso, I. Dabo, Y.-S. Lee, B. Wood, and J. Garg for useful discussions.


\begin{thebibliography}{31}
\bibitem{note_pseudo}
  The core radius of the 3e pseudopotential is
  1.43 $\AA$; the half-bond length at 0 GPa is also 1.43 $\AA$, decreasing to 0.93 $\AA$ at 1200 GPa.
  RRKJ~\cite{PhysRevB.41.1227} pseudowavefunctions closely reproduce all-electron wavefunctions even for
  radii smaller than the matching radius, partially explaining the good results
obtained with the 3e pseudopotential.
\end{thebibliography}
\end{document}